# Flat Bands and Mechanical Deformation Effects in the Moiré Superlattice of MoS$_2$-WSe$_2$ Heterobilayers

Dacen Waters,[1*] Yifan Nie,[2,3] Felix Lüpke,[1**] Yi Pan,[4,5] Stefan Fölsch,[4] Yu-Chuan Lin,[6] Bhakti Jariwala,[6] Kehao Zhang,[6] Chong Wang[1], Hongyan Lv[1,7], Kyeongjae Cho,[2] Di Xiao[1], Joshua A. Robinson,[6] and Randall M. Feenstra[1]

[1]Department of Physics, Carnegie Mellon University, Pittsburgh, PA 15213, USA.

[2]Department of Materials Science and Engineering, The University of Texas at Dallas, Dallas, TX 75080, USA.

[3]Materials Science Division, Lawrence Berkeley National Laboratory, 1 Cyclotron Rd, Berkeley, CA, 94720

[4]Paul-Drude-Institut für Festkörperelektronik, Hausvogteiplatz 5-7, 10117 Berlin, Germany

[5]Center for Spintronics and Quantum Systems, State Key Laboratory for Mechanical Behavior of Materials, Xi'an Jiaotong University, Xi'an, 710049, China.

[6]Department of Materials Science and Engineering, and Center for 2-Dimensional and Layered Materials, The Pennsylvania State University, University Park, PA 16802, USA.

[7]Key Laboratory of Materials Physics, Institute of Solid-State Physics, Chinese Academy of Sciences, Hefei, 230031, China.

*corresponding author: dwaters@andrew.cmu.edu

**present address: Center for Nanophase Materials Sciences, Oak Ridge National Laboratory, Oak Ridge, TN 37831-6487, USA.

## Abstract

It has recently been shown that quantum-confined states can appear in epitaxially grown van der Waals material heterobilayers without a rotational misalignment ($\theta = 0°$), associated with flat bands in the Brillouin zone of the moiré pattern formed due to the lattice mismatch of the two layers. Peaks in the local density of states and confinement in a MoS$_2$/WSe$_2$ system was qualitatively described only considering local stacking arrangements, which cause band edge energies to vary spatially. In this work, we report the presence of large in-plane strain variation across the moiré unit cell of a $\theta = 0°$ MoS$_2$/WSe$_2$ heterobilayer, and show that inclusion of strain variation and out-of-plane displacement in density functional theory calculations greatly improves their agreement with the experimental data. We further explore the role of twist-angle by showing experimental data for a twisted MoS$_2$/WSe$_2$ heterobilayer structure with twist angle of $\theta = 15°$, that exhibits a moiré pattern but no confinement.

## Introduction

Bilayers of two-dimensional (2D) van der Waals material systems have attracted recent attention as a host of a variety of interesting physics and potential for device applications. For example, twisted bilayer graphene was demonstrated to superconduct when the misalignment between the layers is tuned to specific 'magic' angles (*1*). This situation is typified by flat bands in the Brillouin zone of the moiré pattern formed by the two layers (*2*, *3*), which can give rise to correlated behavior such as the superconductivity. This has led to recent interest in so-called 'twistronics', in which tuning of the moiré pattern by changing the misorientation of homobilayers leads to changes in the electronic properties of the system (*4*). The 2D transition metal dichalcogenides have also emerged as interesting systems where flat bands emerge when a small twist angle is introduced in homobilayers of $WSe_2$ (*5*, *6*) or $MoS_2$ (*7*). Another way to realize a moiré pattern, without the need of rotational misalignment, are heterobilayers of materials with nonzero lattice mismatch. Effects of moiré patterns have been studied theoretically for a variety of 2D systems and have revealed, e.g., spatially modulated band gaps (*8*), spin splitting (*9*) in semiconductors, or topological flat bands in trilayer graphene (*10*). In the case of 2D semiconductors, heterobilayers of $MoS_2$ and $WSe_2$ were found to exhibit spatially varying band edge energies, attributed to different local stacking arrangements within the moiré unit cell (*11*). In addition, the present authors recently described the observation of quantum-confined electronic states associated with flat bands in the Brillouin zone of the moiré pattern found in the same system (*12*).

Besides variation of local stacking arrangements, it is expected that the strain in heterolayers will vary spatially as the layers try to match their lattices to correspond to the lowest energy stacking arrangement. This sort of spatially varying strain has been experimentally observed for heterobilayers of graphene on hBN (*13*). Theoretical descriptions of the strain variation for this system qualitatively agree with the experiment (*14–16*). It is well known that strain has a significant impact on electronic structure such as the band gap opening in graphene/hBN bilayers (*17*) or bandgap tuning in the 2D semiconductors (*14–16*). The interface of $WSe_2$-$MoS_2$ lateral heterojunctions were shown to have significant strains (>1%), which lead to the local band gap decrease of ~0.5 eV (*21*). Enhanced photoluminescence intensity at the interface of the same heterojunction has also been observed for this system (*22*), raising questions for the potential of unique spin/valley transport or exciton properties. In addition, out-of-plane deformations are expected in bilayer systems, such that there is minimization of the combined intralayer and interlayer energy (*16*). It is then natural to ask to what degree mechanical deformations play a role in the electronic structure of semiconducting vertical heterobilayers.

The presence of spatially localized quantum-confined states in epitaxially grown $MoS_2$/$WSe_2$ heterobilayers (with no rotational misalignment) were found to be qualitatively explained by local extrema of the conduction or valence band edges, i.e. the states are derived from the lowest (highest) lying conduction (valence) band edge (*12*). In this work, we show that this model holds up well by examining the impact of a twist angle introduced in an exfoliated heterobilayer system, such that a moiré pattern is observed, but no confinement, due to a relatively large band dispersion when there is no spatial variation of band edges. Density functional theory (DFT) calculations considering only local stacking arrangements described well the valence band ordering in the rotationally aligned heterobilayer. However, the conduction band was not well described and requires the inclusion of additional mechanisms to explain the observed confinement. In this work, we show that in-plane as well as out-of-plane deformations have a significant impact on the electronic structure of the bilayer system and explain the observed spatial locations of the confined states. To do so, we combine experimental scanning tunneling microscopy/spectroscopy (STM/STS) results and a detailed DFT study. We find that the strain in the MoS$_2$ varies across the moiré unit cell and that, when this strain variation is included in DFT band structure calculations, a greatly improved agreement is achieved between the predicted and experimentally observed locations of quantum-confined states. In addition, we show that the valence band edge of the present heterobilayer system is sensitive to interlayer separation, from which we infer a non-zero corrugation of the underlying WSe$_2$. The reported behavior is of general applicability for bilayer systems.

## Results and Discussion

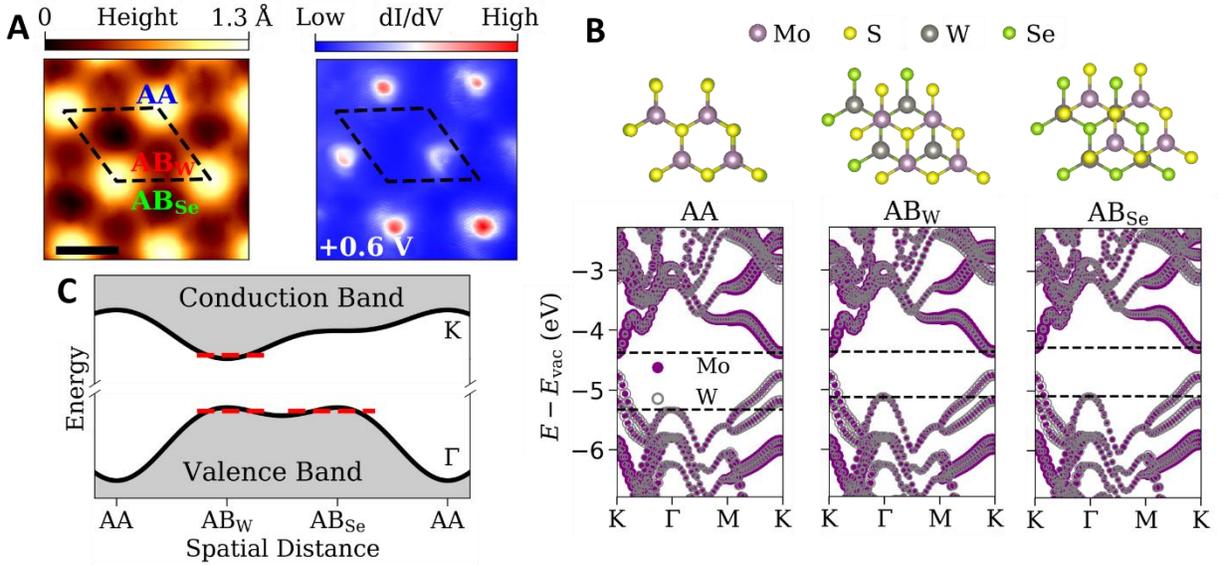

**Fig. 1. Quantum-confined states in MoS$_2$/WSe$_2$ heterobilayer with twist angle of 0°. (A)** STM topography (left) and conductance map (right) of the same area. The moiré unit cell is shown by the black dashed line in both panels. The height map shows the three special registries, denoted by AA for the corrugation maxima and by AB$_W$, and AB$_{Se}$ for the two types of corrugation minima. Sample bias $V_S = +1.5\ V$, scale bar is 5 nm. The conductance map shows the signature of the quantum-confined states at just the AB$_W$ location at $V_S = +0.6\ V$, corresponding to the K point conduction band edge at the AB$_W$ locations. **(B)** Atomic models of AA, AB$_W$, and AB$_{Se}$ stacking configurations (viewed from above) and their corresponding band structures from DFT calculations. Size and color of the data points indicate character of the states. The dashed black lines mark the conduction band minimum at K and the valence band maximum at Γ for each registry (our STS measurements are not sensitive to states at the valence band K point *(12)*). Calculations shown here are done for a lattice constant equal to that of the average of the two materials' relaxed lattice constants ($a = 3.22$ Å). **(C)** Schematic showing the variation of the conduction (K) and valence (Γ) bands and spatial locations of quantum-confined states (indicated by the dashed red lines) as seen in experiment.

**Spatial variation of band edges, neglecting strain**

Heterobilayers of MoS$_2$ and WSe$_2$ were grown on epitaxial graphene and studied using STM/STS (for detail on the sample preparation see Ref. *(12)*). Figure 1A shows the topography (left) and conductance map (right) of the same area. The observed moiré pattern arises from the 3.7% lattice mismatch between the MoS$_2$ ($a = 3.16$ Å) and the WSe$_2$ ($a = 3.28$ Å), which are rotationally aligned in this epitaxially grown sample *(8,9)*. The moiré unit cell is shown by the black dashed line and is found to have a period of $(8.7 \pm 0.2)$ nm and a corrugation of 1.3 Å (at +1.5 V sample bias). The three labeled extrema in the corrugation of the moiré pattern are associated with three different atomic registries *(8,9)*, as shown in Fig. 1B. Spatial locations where the Mo atoms are directly on top of the W atoms (denoted as AA stacking) correspond to corrugation maxima. Alternatively, locations where the metal atom of one layer is opposite the chalcogen atoms of the other layer correspond to the two types of corrugation minima. These are denoted by $AB_W$ ($AB_{Se}$) for the case where the W (Se) atom is visible when looking down through the top layer. Using STS at a temperature of 5 K, it has been shown that states associated with the K point of the conduction band (denoted $K_M$ due to large $MoS_2$ character (*11*)) were confined at the $AB_W$ locations (*12*). Similarly, states associated with the Γ point in the valence band (denoted $Γ_W$ due to large $WSe_2$ character) were found to be confined at both the $AB_W$ and $AB_{Se}$ locations. It should be noted that our measurements are not sensitive to states at the K point of the valence band, and thus our spectra do not reveal the expected confinement associated with those states (*12*).

Neglecting any mechanical deformations of the layers arising from the moiré pattern, the different atomic registries dictate the spatial dependence of the electronic structure. Fig. 1B shows DFT band structure calculations for each registry, following the method of Ref. (*11*), done in a $1 \times 1$ unit cell with the two lattices matched to the average of their relaxed lattice constants ($a = 3.22$ Å) since the moiré unit cell is too large to compute in its entirety. In a semi-classical picture, the spatially varying band edges can be treated as a potential with the periodicity of the moiré pattern. Wherever there is a local maximum (minimum) in the valence (conduction) band, there is an effective potential-well such that there is confinement at that spatial location (Fig. 1C). These confined states then manifest as a sharp peak in the local density of states near the band edge (i.e. a flat band in the moiré pattern Brillouin zone), which are evident at the $AB_W$ locations in the conductance map shown in Fig. 1A. The potential can be written as

$$V(\vec{r}) = \sum_{\vec{G}} V_{\vec{G}} e^{i\vec{G}\cdot\vec{r}} \qquad (1)$$

where $\vec{G}$ are reciprocal lattice vectors of the moiré pattern. Following the method of Ref. (*17*), assuming a slow spatial variation, we perform a Fourier decomposition of the band edge energies and keep only the zero and first order terms of the potential (see Supplemental Material for details). The resulting band structure depends on the amplitude of the variation, $V_G$, and the moiré wavelength, $L$. As we will show below, for a large variation of the band edge energies, the lowest lying energy band flattens such that a sharp peak is found in the density of states.

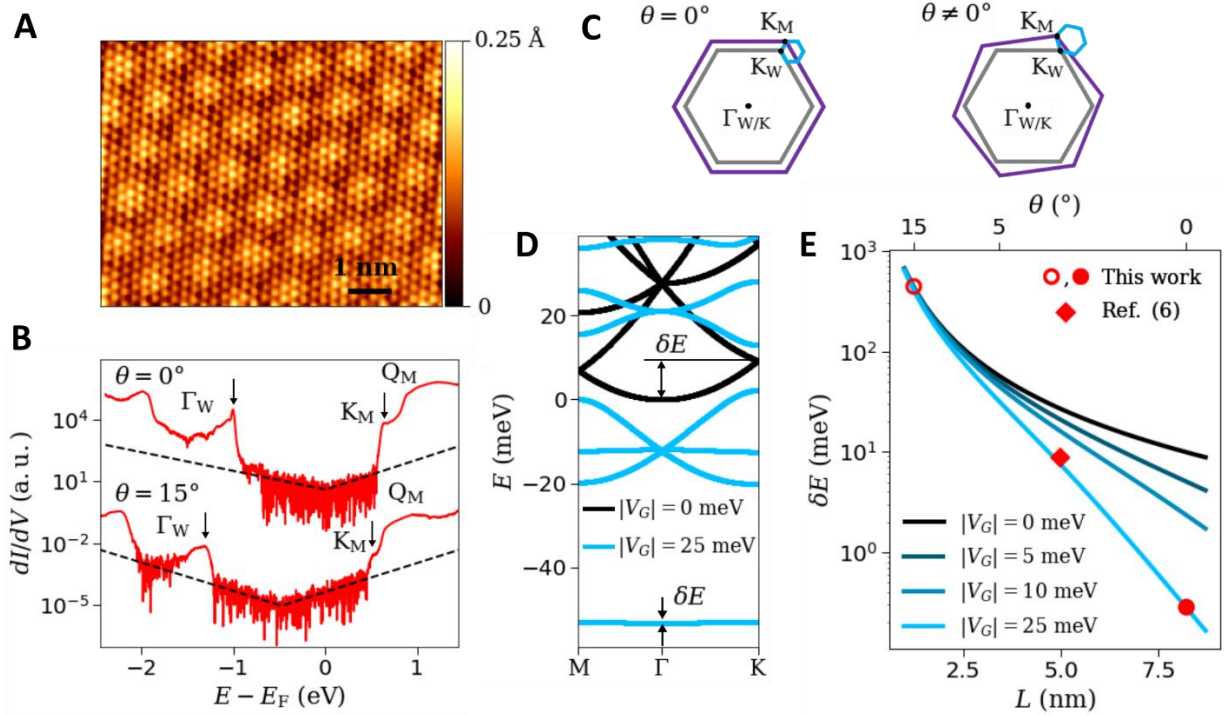

**Fig. 2. Twisted MoS$_2$/WSe$_2$ heterobilayer.** **(A)** Topography of an exfoliated MoS$_2$/WSe$_2$ heterobilayer structure with a twist angle of $\theta = 15°$. Simultaneous atomic resolution and the moiré pattern ($L \approx 1.2$ nm) are evident in the topograph. **(B)** Spectra for the grown heterobilayer ($\theta = 0°$) at an AB$_W$ location and the exfoliated heterobilayer ($\theta = 15°$) at a corrugation minimum, showing similar features, expect with an overall shift in energy which can be explained to be due to substrate effects. Features associated with parts of the single layer Brillouin zones are indicated. Notably, for $\theta = 0°$ sharp peaks are observed (particularly for Γ$_W$), associated with flat bands in the moiré Brillouin zone, which are not evident for the $\theta = 15°$ case, indicated by the arrows. Black dashed lines show a voltage-dependent noise level (see Methods for details). **(C)** Schematic representation of the Brillouin zone for the two layers with no twist angle (left) and a small non-zero twist angle (right). The purple (grey) hexagon represents the MoS$_2$ (WSe$_2$) Brillouin zone. The blue hexagon is the Brillouin zone of the moiré pattern formed. **(D)** Band structure for the Brillouin zone of the moiré pattern with $\theta = 0°$. $|V_G|$ is the magnitude of the spatial variation of the band edge. With $|V_G| = 0$, the dispersion of the lowest lying energy band, $\delta E$, is on the order of 10 meV. With $|V_G| = 25$ meV and $\theta = 0°$, the dispersion of the lowest lying energy band is <1 meV. **(E)** Dispersion of the lowest lying energy band, $\delta E$, as a function of moiré wavelength, $L$, or twist angle, $\theta$, for various values of $|V_G|$. The dispersion is very small for a large value of $|V_G|$ and long moiré wavelength (or small twist angle), indicating that a flat band and confinement will occur. For short moiré wavelengths (or large twist angles), the dispersion is large regardless of the value of $|V_G|$. Data points from experiment from this work and Ref. (6) shown. The filled circle indicates that flat bands are observed in this work for $\theta = 0°$, and the open circle represents the experimental results for $\theta = 15°$ where flat bands are not observed. Ref. (6) reports the observation of flat bands associated with the Γ point in the valence band for a twisted homobilayers of WSe$_2$ for $\theta = 3°$.

In Fig. 2A, we show an STM image of a heterobilayer of exfoliated MoS$_2$ on WSe$_2$ (on a graphite substrate) with $\theta \approx 15°$ (directly measured from separate atomic resolution images, see Supplemental Material for details). A small wavelength moiré pattern is formed, $L \approx 1.2$ nm. The moiré wavelength is consistent with the expected periodicity given by $L = a/\sqrt{\delta^2 + \theta^2}$, where $a$ is the lattice constant of the MoS$_2$ (or WSe$_2$, whichever is used amounts to a small difference in the calculated moiré wavelength), $\delta = 3.7\%$ is the lattice mismatch, and $\theta$ is the twist angle in radians. A spectrum from an AB$_W$ site on the grown heterobilayer ($\theta = 0°$) is shown in Fig. 2B with sharp peaks associated with the confined states indicated by arrows. For the small wavelength moiré pattern ($\theta = 15°$), the same spectroscopic features are observed (with an overall shift due to different substrates), but with no sharp peaks associated with quantum-confinement. Confinement is not observed at any spatial location for the twisted heterobilayer (see Supplemental Material). The lack of confinement can be explained by the potential given by Equation 1. In Fig. 2C, we show a schematic representing the Brillouin zone formed when there is a small lattice mismatch with no twist angle (left) and with a small non-zero twist angle (right). A moiré pattern forms in both scenarios but will have a shorter wavelength (larger Brillouin zone) when a small twist angle is present, as seen in Fig. 2A. Fig. 2D shows the band structure for $\theta = 0°$ with no band edge variation, $|V_G| = 0$, and for a band edge variation corresponding to that of the $\Gamma_W$ band edge in the grown heterobilayer, $|V_G| = 25$ meV. The lowest lying energy band flattens when $|V_G| = 25$ meV such that the dispersion of this band is very small ($< 1$ meV), which we attribute to the sharp peak in the density of states observed in the spectra of Fig. 2C for $\theta = 0°$. Fig. 2E shows the dispersion of the lowest lying energy band as a function of moiré wavelength for various values of $|V_G|$. For small moiré wavelengths (or large twist angles), the dispersion increases such that there is no longer a flat band in the moiré pattern Brillouin zone, regardless of the value of the band edge variation. Fig. 2E also shows the experimental data points for flat bands associated with the $\Gamma_W$ point, with filled (open) data points indicating where confinement and flat bands are observed (not observed). Ref. (*6*) reports flat bands in a twisted homobilayer of WSe$_2$ with $\theta = 3°$ associated with the Γ point in the valence band (i.e. $\Gamma_W$). This compares well with our data and the prediction from our model, that a flat band with a dispersion of <10 meV emerges for their reported valence band edge variation. The mechanism here is generally applicable to semiconductor hetero- or homo-bilayers when a moiré pattern is evident. In addition, interesting correlated phases might emerge when the flat bands observed in this system are gated to the Fermi level such as in twisted homobilayers of WSe$_2$ (*5*).

Returning to the grown heterobilayer and the observed band edge variation, we must consider the ordering of band edge energies to explain the observed confinement. Dashed lines in the band structures of Fig. 1B mark the edges of the $\Gamma_W$ valence band and the $K_M$ conduction band. In the former case, we see that the $\Gamma_W$ valence band has nearly the same energy for the AB$_{Se}$ and AB$_W$ registries, with the band edge for the AA registry being 0.25 eV lower. Hence, we find that the ordering of the band edges aligns well with the experimentally observed locations of the valence-band confined states as shown in Fig. 1C. However, for the $K_M$ conduction band, the band structures do not accurately predict where confinement is observed. Specifically, we find from the computed band structure nearly equal $K_M$ band edge energies for the AA and AB$_W$ registries, with that for the AB$_{Se}$ registry being only 100 meV higher. However, in experiment, the $K_M$ conduction band edge is found to be localized only for the AB$_W$ registry (Fig. 1C). The band structure calculations, however, do not consider any of the effects from mechanical deformations. In the following, we explore in detail the effects of spatially varying strain and interlayer separation on the electronic structure of this system.

**Effects of strain variation on electronic structure**

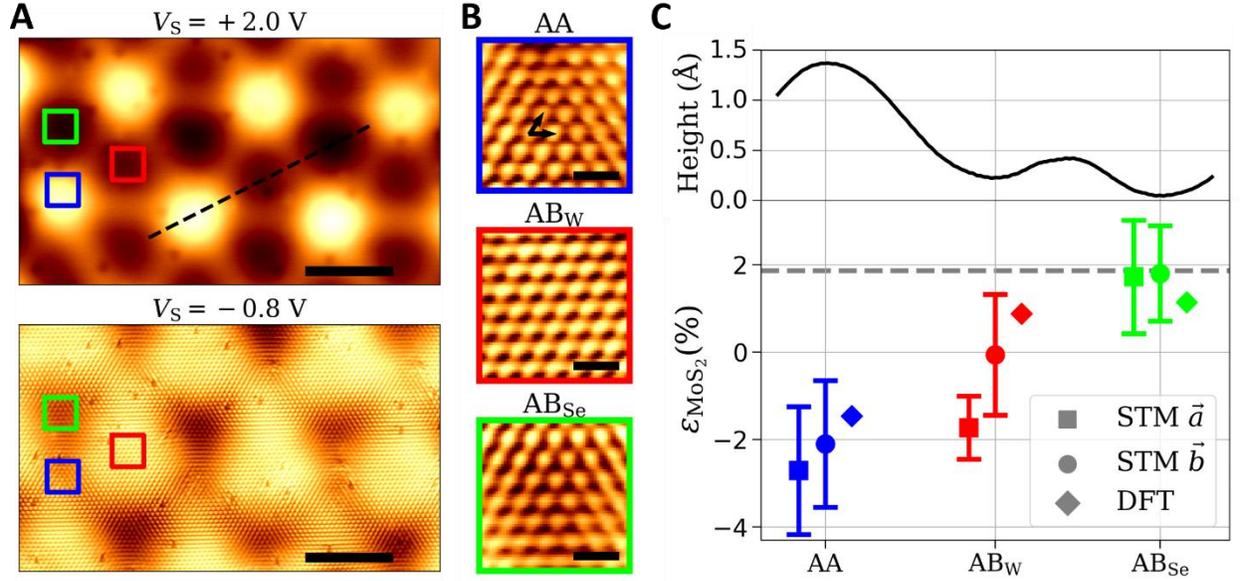

**Fig. 3. Spatially varying strain in the moiré pattern. (A)** STM images of the same area at different sample biases for the grown heterobilayer ($\theta = 0°$). The top panel shows an image at $V_S = +2.0\ V$ sample bias, where only the moiré period is observed. The bottom panel shows an image at $V_S = -0.8\ V$, where both the moiré period and atomic resolution of the MoS$_2$ is observed. Representative AA, AB$_W$, and AB$_{Se}$ locations are shown by blue, red, and green boxes, respectively. Scale bars are 5 nm. **(B)** Zoomed-in atomic resolution images for the regions marked in (A). $V_S = -0.8\ V$. Scale bars are 5 Å. From these images, the in-plane strain of the MoS$_2$ is directly measured. Lattice vectors used to extract the strain are shown in the AA panel with $\vec{a}$ pointing up and to the right and $\vec{b}$ pointing horizontal. **(C)** Height profile (top panel) taken along the dashed line in (A). Strain in the MoS$_2$ at each position (bottom panel, data points for the same locations are offset laterally for clarity). The dashed line shows the strain associated with the average lattice constant of the two materials ($a = 3.22$ Å). Error bars show the standard deviations for a series of measurements from each type of location visible in the $V_S = -0.8\ V$ image in panel (A). Squares (circles) show the strain measured by comparing the magnitude of $\vec{a}$ ($\vec{b}$) to the nominal lattice constant of MoS$_2$. Diamonds show the predicted strain values from DFT calculations at each registry.

Figure 3A shows two large scale STM images of the same area at different tunneling voltages. The $V_S = +2\ V$ image shows many unit cells of the familiar moiré pattern, with representative areas of AA, AB$_W$, and AB$_{Se}$ locations highlighted by the color-coded squares. The same areas are indicated in the $V_S = -0.8\ V$ image. At these latter tunneling conditions, we can achieve simultaneous resolution of the moiré pattern as well as atomic resolution of the MoS$_2$ layer. Zoomed-in images of the corresponding areas from $V_S = -0.8\ V$ are shown in Fig. 2B, from which we extract the strain in the MoS$_2$ at each location, by measuring the local lattice constant (see Supplemental Material for details). The results for each location are plotted in Fig. 3C, along with a height profile taken from the $V_S = +2\ V$ image. The grey dashed line in the bottom panel of Fig. 3C indicates the strain corresponding to the average of the lattice constants of the MoS$_2$ and WSe$_2$ ($a = 3.22$ Å). We see that at the AB$_{Se}$ location, where the corrugation is lowest such that the layers are presumably closest together, there is a relatively large tensile strain due to the layers trying to match their lattice constants. At the AA location, where the layers are furthest apart, the MoS$_2$ appears to be 'compensating' for the deformation at the AB$_{Se}$ location by relaxing past the nominal MoS$_2$ lattice constant. The strain variation in the MoS$_2$ is measured to be $\Delta\varepsilon_{MoS_2} = \varepsilon_{MoS_2}(AB_{Se}) - \varepsilon_{MoS_2}(AA) = 4.17 \pm 1.45\%$, which is relatively large in comparison to other bilayer systems such as graphene/hBN which has been reported to have a strain variation of ~2% (*13*). In addition, the strain measured at the AB$_W$ corrugation minimum is compressive, where one would expect a tensile strain. However, it has been found that electrostatic forces due to the presence of the STM tip can distort 2D materials such as graphene (*23–26*), influencing the observed lattice constant and thus enhancing the measured strain signal. Since the atomic corrugations only become apparent for smaller negative voltages, the STM tip must be very close to the sample to maintain a measurable current, and therefore large electrostatic forces are likely present. Lateral force between the tip and sample might also explain why we see a difference in the observed strain measured along the two axes at the AB$_W$ location, instead of a uniform biaxial strain as seen at the other locations, and evidence for such a distorting effect is presented in the Supplementary Material.

Domain formation and relaxation in bilayer systems have been studied theoretically in detail recently. Following the method of ref. (*16*), we assume the interlayer potential energy can be obtained by summing over local energies of different registries. The total energy, consisting of interlayer potential energy and intralayer elastic energy, is then minimized in order to find the relaxed configuration of the moiré superstructure (see Supplemental Material for details). The theoretically calculated strain at each location are shown in Fig. 2C. Overall, we find very good agreement with the results from STM topography. The strains for the AA location and AB$_{Se}$ registry agree very well with the experimental results, with the predicted strain variation, $\Delta\varepsilon_{MoS_2} = \varepsilon_{MoS_2}(AB_{Se}) - \varepsilon_{MoS_2}(AA) = 2.60\%$, being slightly smaller than the experimentally observed value. The predicted strain at the AB$_W$ registry is more in line with what one would expect for a corrugation minimum, as compared to the experimental result mentioned above. We will show next that the predicted strain values produce significant improvement for the agreement between theoretical band edge positions and STS observations. In addition, we expect that there will be strain variation in the bottom layer of the bilayer, which cannot be measured by our surface probe technique, but the theoretical results predict that a similar strain variation will occur in the WSe$_2$. Therefore, we use the DFT-predicted strain in the following analysis for both layers.

To explore the effects of strain variation on the electronic structure, we perform DFT calculations of the heterobilayer for each registry, but for different values of the matched lattice constant, i.e. including the strain. The results for the AA location are shown in Fig. 4A, with the energies of the conduction band minima at the K and Q points marked by dashed black lines. The effects on the band structure arising from the strain are manifold. First, the band gap shrinks as the MoS$_2$ is subjected to increasing tensile strain, due to the conduction band at the K point being driven downwards. Second, for large compressive strains, the ordering of the conduction band minimum at the K and Q points invert, causing this stacking arrangement to have an indirect gap. In addition, we see that the valence band experiences a shift, although to a lesser degree; the valence band edge at the Γ point is driven up, while there is little change elsewhere in the Brillouin zone. We can quantify the effect of strain at each point in the Brillouin zone by extracting the band edge energy. The result is shown in Fig. 4B for the K and Q points (K$_M$ and Q$_M$, respectively) for the AB$_W$ and AB$_{Se}$ registries. We fit a linear behavior to the extracted band edge energies, with the slope of the fit being the deformation potential. From this point on, we assume that the band edge shifts in the conduction band are associated only with the strain in the MoS$_2$, and any shifts in the valence band are associated only with the strain in the WSe$_2$. This assumption is motivated by the fact that the deformation potentials in the conduction (valence) band are very similar to those obtained in calculations for individual monolayers of MoS$_2$ (WSe$_2$), as demonstrated in the Supplemental Material.

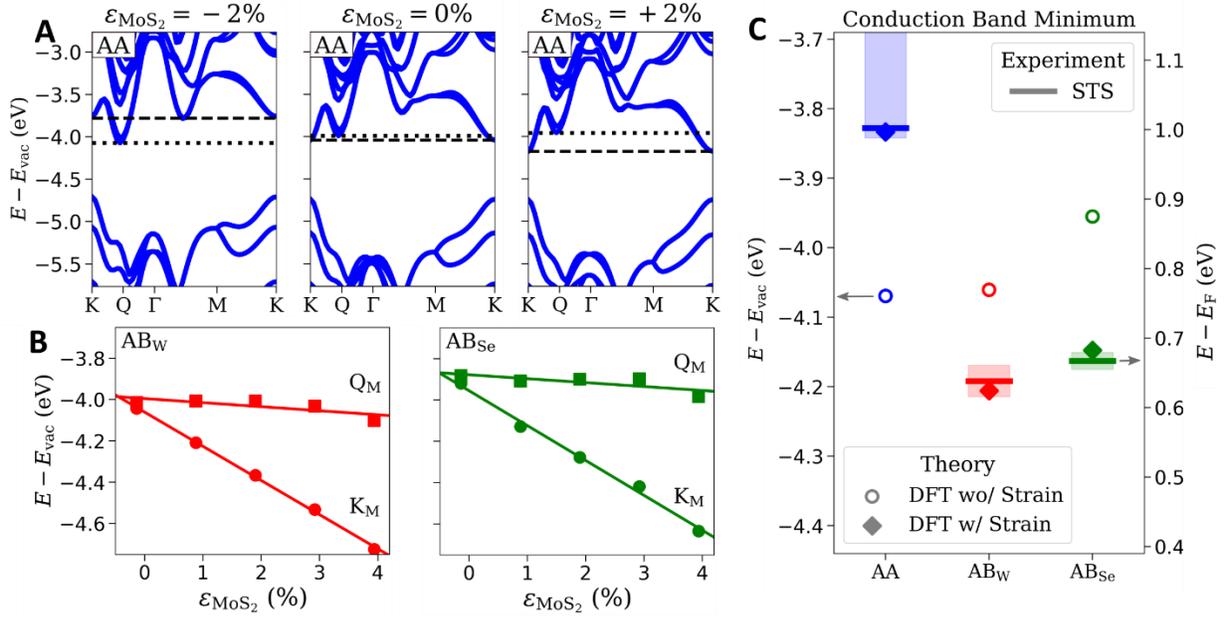

**Fig. 4. Effects of in-plane strain on the conduction band edge. (A)** Band structure calculations for the AA registry for different values of strain in the MoS$_2$. The K$_M$ and Q$_M$ band edges are marked by the black dashed and dotted lines, respectively. **(B)** Band edge energies from DFT for the other two registries as a function of strain. Solid lines are the linear best fit to extract the deformation potential. See text for more details. **(C)** Theoretically predicted and experimentally observed K$_M$ band edge energies for each registry. Left axis shows the DFT results without strain (i.e. the same as in Fig. 1B) as open circles and the DFT results incorporating strain variation as solid diamonds. Right axis indicates the band edge energies observed in STS and associated errors (horizontal thick lines, with shaded regions; see Supplemental Material for details). At the AA locations, we don't observe a K$_M$ band edge in STS, just the Q$_M$ band edge. The observed band edge position is therefore a lower limit on the actual location of the K$_M$ band edge at the AA location, as indicated by the shaded region extending all the way up.

The deformation potential for the K$_M$ band edge for the AA registry is found to be $-0.16$ eV/%. Combining this with the large compressive strain predicted for the AA location ($\varepsilon_{MoS_2}(AA) = -1.46\%$) causes the K$_M$ band edge at the AA location to be driven up by 0.23 eV, while the band edges of the AB$_W$ and AB$_{Se}$ registries are pushed down (Fig. 4C). Therefore, the quantum-confined states are no longer predicted to occur at the AA location, in contrast to the prediction from the 'zero-strain' DFT results. The quantum-confined states are predicted to occur at the AB$_W$ locations, which is exactly what is seen in experiment (Fig. 1A). Accounting for strain variation in the MoS$_2$ greatly improves the agreement between experiment and theory, as can be seen in Fig. 4C, where we show band edge energies extracted from spectra taken at each location. At the AA locations, we don't observe a K$_M$ band edge, which is successfully explained by incorporating the predicted strain variation: for large compressive strains, the AA conduction band minima occurs at Q rather than K (Fig. 4A). Therefore, we expect tunneling at the conduction band edge to be dominated by Q$_M$ rather than K$_M$ (see Supplemental Material for details).

### Interlayer separation dependence

Let us now turn our attention to out-of-plane deformations. In our STM measurements, we are only able to measure height differences of the top-most layer, with these heights being in general a convolution of the sample morphology and its electronic structure. However, at large positive sample voltages, local differences in the electronic structure of this system are small such that the measured topography is close to the actual morphology of the sample (see Supplemental Material). DFT calculations predict an optimal interlayer separation of 6.9 Å, 6.29 Å, and 6.30 Å for the AA, $AB_W$, and $AB_{Se}$ locations, respectively. Therefore, the corrugation predicted by DFT is 0.6 Å, less than half of the observed ~1.3 Å corrugation from the $V_S = +2$ V topography such as that in Fig. 2C. It is qualitatively easy to rectify this discrepancy, noting that strain variation will induce buckling of the layers in order to minimize the system's energy (*14*). This buckling will then change the interlayer separation, deviating from the optimal separations predicted by DFT. To analyze this situation further, we study the impact of interlayer separation on the electronic structure of the bilayer system.

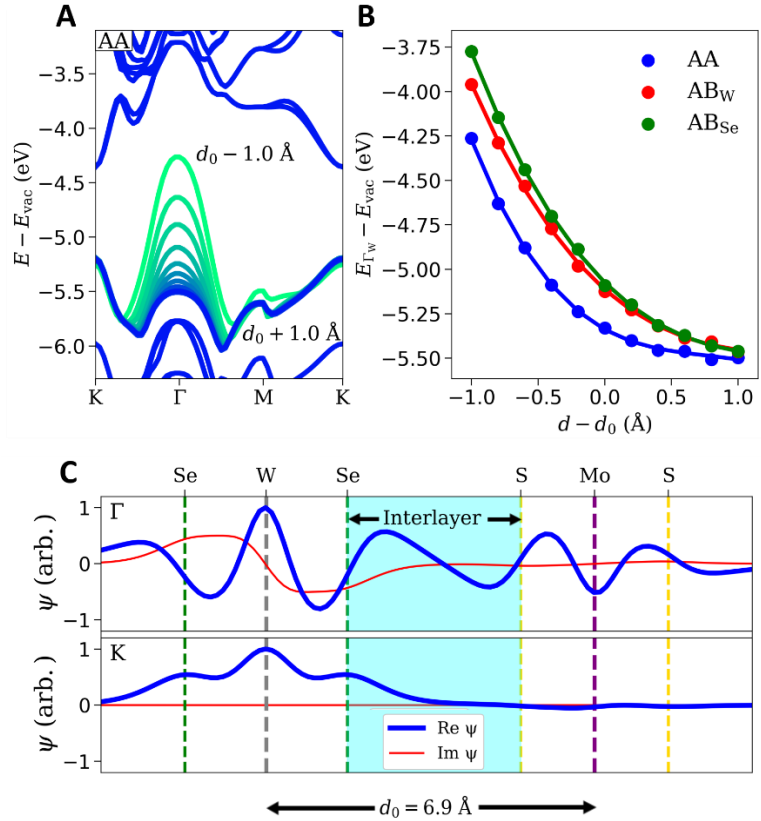

**Fig. 5. Interlayer separation dependence. (A)** Band structure for the AA registry at a fixed lattice constant ($a = 3.22$ Å), but varying interlayer separation. For clarity, only the $d_0 + 1.0$ Å calculation is shown for bands other than the valence band (variation everywhere else is very small, <50 meV, comparable to the variation shown in the valence band at K or M). **(B)** $\Gamma_W$ valence band edge energies extracted for each registry as a function of interlayer separation. The optimal interlayer separation (vertical distance between the planes of metal atoms in each layer) for the AA, $AB_W$, and $AB_{Se}$ locations are $d_0 = 6.9$ Å, 6.29 Å, and 6.30 Å, respectively. The solid line shows the best fit to a third order polynomial. **(C)** Wavefunctions for the AA registry valence band at the Γ (top) and K (bottom) points, for a representative location in the 1x1 unit cell. Vertical dashed lines denote the plane of the designated atoms. Interlayer region is highlighted in blue.

Fig. 5A shows the calculated band structure for the AA registry as the interlayer separation is varied, but with a fixed lattice constant. As the layers are pushed together, we see that the valence band edge at the Γ point is driven up, while other points in the Brillouin zone are mostly unaffected. In contrast, as the layers are pulled apart, changes in the valence band are much smaller since the layers are becoming more like free standing monolayers. This behavior is shown for each registry in Fig. 5B, where the band edge positions are plotted as a function of interlayer spacing as compared to each registry's optimal spacing. The dependence on interlayer separation can be attributed to a large interlayer character of the wave function at the Γ point. In Fig. 5C, we show the wave function (*27*) at a representative location in the $1 \times 1$ unit cell calculation for both the Γ (top) and K (bottom) points in the valence band. It is easily seen that at the Γ point, the wave function has a relatively large amplitude in the interlayer region (despite having a node between the two layers). However, the wave function at the other points in the Brillouin zone, such as the K point shown, is localized around the $WSe_2$ layer and decays more quickly in the interlayer region.

Let us consider how this dependence of the $Γ_W$ energy on the interlayer separation will affect the location of the confined states in the valence band (and how possible corrugation of the bottom layer of the heterobilayer, the $WSe_2$, might play a role). The 'zero-strain' DFT results shown in Fig. 1, which well predict the valence band edges and thus the locations of the confined states, assumes the equilibrium spacing of the two layers. However, the layers cannot be at their equilibrium separation because the predicted corrugation amplitude is less than the observed one (Fig. 3C) by a factor of two. If the layers are not at their equilibrium separation, then there will be an effect on the band structure, particularly for the $Γ_W$ valence band edge. Therefore, we conclude that there is *both* in-plane strain and out-of-plane displacement of the underlying $WSe_2$ layer. Utilizing our DFT results, we can model the band edge positions of this system as a function of the strain and interlayer separation (for details see Supplemental Material). Using the experimentally observed valence band edge energies, corrugation, and step height from $WSe_2$ to $MoS_2$ as an input to the model, we infer the corrugation of the $WSe_2$ layer. The results are shown in Fig. 6, where we find that the corrugation of the underlying $WSe_2$ is 70% of the corrugation in the $MoS_2$ and that it is in phase with the $MoS_2$ corrugation. The magnitude of the inferred corrugation in the $WSe_2$ is sensitive to the measured step height from the $WSe_2$ to $MoS_2$, which we discuss in the Supplemental Material.

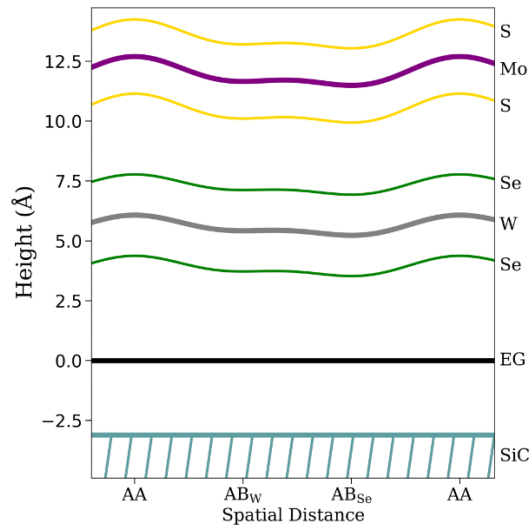

**Fig. 6. Inferred corrugation of WSe₂.** Height profiles of $MoS_2$ and $WSe_2$ determined by combining theoretical DFT predictions and experimentally observed signals from STM (see text and Supplemental Material for more details). Atomic species for each layer designated on the right, as well as the underlying epitaxial graphene (EG) and SiC

substrate. Separation distances of WSe$_2$-EG (at the AB$_{Se}$ registry) and EG-SiC substrate are assumed to be 5.2 Å *(28)* and 3.1 Å *(29)*, respectively.

## Conclusion

We have shown that in-plane strain variation and out-of-plane deformations have a significant impact on the electronic properties of epitaxially grown MoS$_2$/WSe$_2$ heterobilayers. These effects are quite general and extend to other 2D systems, including twisted bilayers. These results are expected to have significant impact on the optoelectronic properties of semiconductor heterostructures, the details of which should be explored in detail. Exciton behavior will be greatly affected, particularly if the direct gap nature of the heterostructure is modulated spatially, as is predicted from our results. Additionally, our results demonstrate that strain can be used as a powerful tuning parameter to engineer the electronic structure of van der Waals multilayers. Deformation-dependent electronic structure will have a major impact on optoelectronic studies, theoretical predictions, and the behavior of devices and must be studied in detail to gain a complete understanding of a given bilayer structure. Lastly, we note that MoS$_2$/WSe$_2$ heterobilayers could exhibit exotic electronic phases if the experimentally observed flat bands are gated to the Fermi level.

## Acknowledgements

Funding: This work was supported in part by the A. von Humboldt Foundation and by the Center for Low-Energy Systems Technology (LEAST), one of six centers of STARnet, a Semiconductor Research Corporation program sponsored by Microelectronics Advanced Research Corporation (MARCO) and Defense Advanced Research Projects Agency (DARPA). F.L. and D.W. were supported by NSF DMR-1809145 for the STM measurements of the twisted heterobilayer. The authors gratefully acknowledge NSF DMR-1626099 for acquisition of the STM instrument. Y. Pan also acknowledges the support from the National Key R&D Program of China (2017YFA0206202) and the National Science Foundation of China (11704303). Y.N. acknowledges Texas Advanced Computing Center (TACC) for providing computation resources. The Moiré calculation (C.W. and D.X.) is supported by the Department of Energy, Basic Energy Sciences, Grant No. DE-SC0012509. H.L. acknowledges the financial support from the China Scholarship Council (No. 201904910165) during the visit at Carnegie Mellon University.

Author contributions: D.W., F.L, Y.N., S.F, and R.F. performed STM measurements and data analysis. Y.N. and K.C. performed DFT calculations to determine deformation potentials. C.W. H.L, and D.X. performed theoretical calculations to incorporate relaxation in the moiré unit cell. Y.-C. L. B.J., K.Z. and J.R. performed the growth of the epitaxial heterobilayer. D.W. and F.L. created the twisted heterobilayer and performed the analysis. D.W. and R.F. prepared the manuscript, with contributions from all coauthors.

Competing interest: The authors declare they have no competing interests.

## Materials and Methods

The metal-organic chemical vapor deposition (MOCVD) growth of WSe$_2$ is performed at 700 Torr using H$_2$ as a carrier gas at 800°C, with W(CO)$_6$ and H$_2$Se precursors being introduced separately into a cold wall vertical reactor chamber and their respective flow rates controlled via mass flow controllers. The optimized condition for the growth was based on a recent detailed study of WSe$_2$ growth (*30*). On top of these layers MoS$_2$ is deposited by CVD, using 2 mg MoO$_3$ and 200 mg sulfur powder as the optimal precursor ratio for synthesis performed at 850°C. The substrate consists of epitaxial graphene formed on SiC. The twisted heterobilayer fabrication followed the method described in Ref. (*31*).

The STM/STS measurements were carried out in ultrahigh vacuum at 5 K. Electrochemically etched tungsten tips cleaned in UHV by Ne ion bombardment and/or electron beam heating were used. STM images were recorded in constant-current mode using currents in the range 0.01 – 0.1 nA; bias voltages refer to the sample with respect to the STM tip at ground potential. STS measurements of the differential tunneling conductance dI/dV were carried out using standard lock-in technique (modulation frequency 675 Hz with a peak-to-peak modulation of 10 mV) to probe the local density of electronic states. STS measurements were performed with a z-ramp on the order of $1-2$ Å/V to increase the dynamic range of the measurement (*32*). A normalization factor of $e^{-2\kappa z}$ is applied to the spectra shown, where $\kappa \sim 1$ Å$^{-1}$ is determined experimentally. The noise level is determined to be one standard deviation above the average of the measured conductance within the band gap of the heterobilayer, with z-dependent normalization applied to get the voltage dependent noise levels indicated in the figures by black dashed lines.

DFT calculations were done using the Vienna Ab-Initio Simulation Package (*33*) with the projector-augmented wave method (*34*), employing the Purdew-Burke-Ernzerhof generalized gradient approximation exchange-correlation functional (*35*) together with dipole corrections obtained by Grimme's DFT-D2 method (*36*). The wave functions are expanded in plane waves with a cutoff energy of 400 eV, and the energy convergence criteria for electronic and ionic optimization are 10$^{-4}$ eV and 0.01 eV/Å, respectively. Integration over the first Brillouin zone is carried out with a Γ-centered 24×24×1 k-point mesh for the wave function calculations. Spin-orbit coupling is included in the calculation of electronic structure. A vacuum region of over 10 Å in the direction normal to the 2D material layers is added to minimize the interaction between the adjacent supercell images. See Supplemental Material for further details. In the moiré calculation, the energy landscape is sampled by shifting the top layer relative to the bottom layer over a 9x9 grid in the unit cell, and the dispersion correction with the optB86b-vdW functional (*37*) is adopted. For each stacking, the in-plane positions of Mo and W atoms are fixed, while the other atoms are fully relaxed. See Supplementary Material for further details.

# Supporting Information

# Flat Bands and Mechanical Deformation Effects in the Moiré Superlattice of MoS$_2$-WSe$_2$ Heterobilayers

**Scanning Tunneling Spectroscopy of Quantum-Confined States**

Detailed spectroscopic results are shown in our previous work (*1*), but here we briefly touch on these results in the context of the mechanical deformation effects and DFT analysis presented in the main text. Figure S1A shows a topography image at $V_S = -1.5$ V as well as a line along which spectra were taken. The color-coded spectra are shown in Fig. S1B, with spectroscopic features marked, including some defect states evident near the valence band edge. The marked band edges are identified with those in previous work, Refs. (*1*, *2*), and are plotted as a function of spatial distance in Fig. S1C, from which the spatial variation of the band edge positions is obvious. For the conduction band edges, the Q$_M$ and K$_M$ band edges were identified by an inflection point in the spectrum, except for the K$_M$ band edge at the AB$_{Se}$ locations, where the peak associated with the confined state identifies the band edge. For the valence band, the Γ$_W$ band edge is identified by the peak in the spectra, as well as the lower lying valence band edges previously identified as a combination of multiple states (*2*). All spectra were taken at 5 K with a *peak-to- peak* modulation voltage of $V_{\mathrm{mod}} = 10$ mV (which is mistakenly reported in the main text of Ref. (*1*) to be the RMS value).

This revision to the modulation voltage (with the corrected value corresponding to an RMS voltage of 3.53 mV) has some impact on the discussion of Ref. (*1*). With this reduced RMS value, the expected energy resolution of our spectra is $\Delta E = 9$ meV at 5 K, and 26 meV at 80 K. The significant difference between these two values now makes it much easier to understand why, as reported in Ref. (*1*), the sharp band-edge peaks that we observe at specific locations in the moiré unit cell for measurements at 5 K, are found to be absent at 80 K. (Additionally, the observed peak widths of ~25 meV found in the 5 K data are now recognized as being *intrinsic* widths, i.e. not due to the modulation).

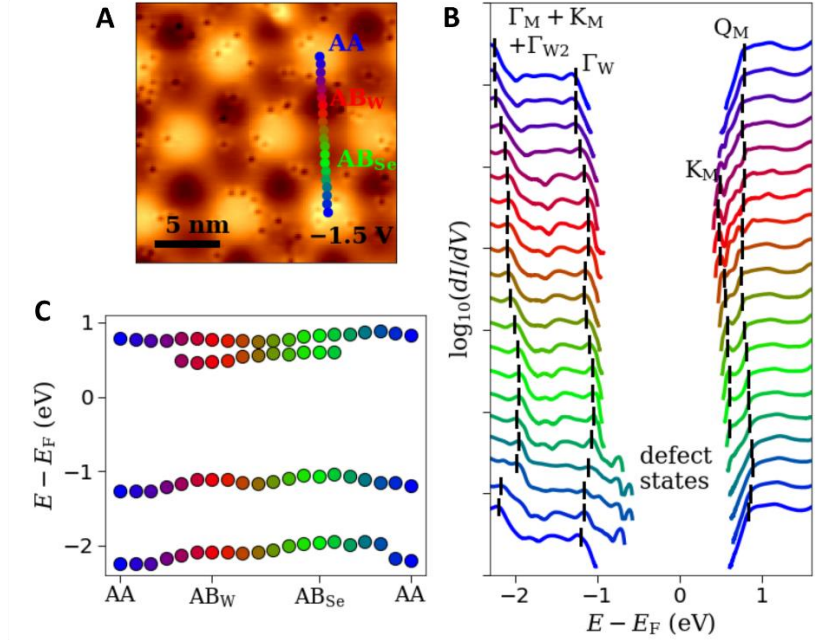

**Fig. S7. Quantum-confined states in MoS$_2$/WSe$_2$ heterobilayer. (A)** STM topography of the moiré pattern with multiple point defects visible. **(B)** STS measurements along the line shown in (A). Several spectroscopic features are marked (see text for details) to illustrate the spatial variation within the moiré pattern. Defect states are evident in the valence band at the lower AA location. **(C)** Spatial mapping of the spectroscopic features marked in (B).

As discussed in the main text, our model incorporating strain in the MoS$_2$ layer results in the order of the K$_M$ and Q$_M$ band edges inverting for large values of compressive strain (Fig. 4A). While this is difficult to confirm explicitly in experiment, it does align well with our scanning tunneling spectroscopy (STS) results. As can be seen in Fig. S1B, the K$_M$ feature (either a peak near the AB$_W$ registry or a shoulder near the AB$_{Se}$ registry) is clearly apparent except at the AA registries. The spectroscopic feature identified as the K$_M$ band edge at the AB$_W$ location seems to shift upwards in energy, merging with the shoulder associated with the Q$_M$ band edge at the AA location. Therefore, we treat the identification of the Q$_M$ band edge as a lower limit on the K$_M$ band edge energy for the AA locations (Fig. 4C), since the two cannot be distinguished without further study.

## Twisted heterobilayer

The twisted heterobilayer was made using the flip-stack method described in Ref. (*3*). Briefly, exfoliated flakes were transferred in reverse order using a polymer film. The assembled stack was then flipped and placed on an evaporated gold pattern which is used as the drain electrode for STM measurements. The sample was then annealed in ultrahigh vacuum to remove the polymer film underneath the stack and then transferred to the STM chamber. The individual exfoliated layers are shown in Fig. S2A. The finished stack is shown in Fig. S2B, with the overlap region where the moiré pattern is observed indicated in bright blue. The small twist angle was achieved by optically aligning the sharp straight edges of the exfoliated flakes during the transfer process. Fig. S2C shows a large scale STM image along with a height profile taken along the indicated line. The small step at $x \approx 0.15$ μm is $\Delta z \approx 0.6$ nm, corresponding to a monolayer of MoS$_2$. The $\Delta z \approx 1.8$ nm step at $x \approx 0.3$ μm is the edge of the bulk portion of the MoS$_2$ flake, while the largest step at $x \approx 0.55$ μm is the MoS$_2$ flake draping over the bulk-to-monolayer step of the underlying WSe$_2$.

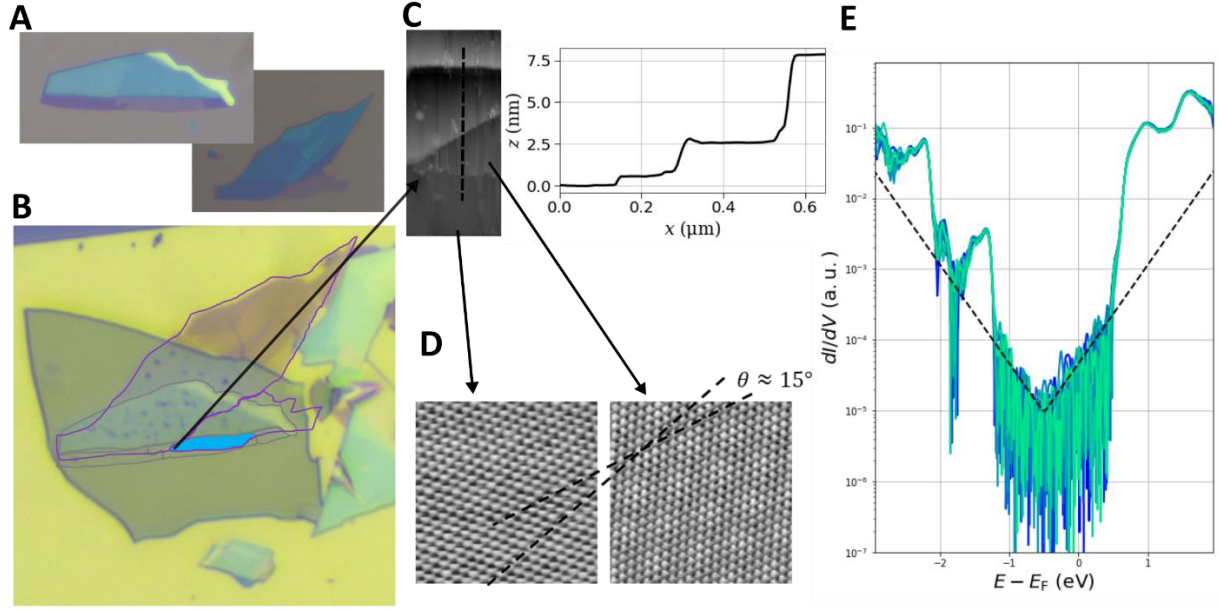

**Fig. S8. Fabrication and characterization of the twisted heterobilayer. (A)** Optical images of the exfoliated WSe₂ (left) and MoS₂ (right). The thin purple regions in each image are monolayers, as confirmed by height profiles taken in STM measurements. **(B)** The assembled heterobilayer. Outlines of the WSe₂ (MoS₂) are shown in grey (purple) to highlight features. The light blue region is where the overlap occurs and the moiré pattern is observed in STM. The large dark-green flake underneath the heterostructure is the graphite substrate. The yellow portion of this image is the evaporated gold pattern used as the tunneling drain during measurements. The arrow indicates that the large scale STM image is taken from that area. **(C)** Large scale STM image with a height profile along the dashed line shown. The first step at $x \approx 0.15\ \mu m$ where $\Delta z \approx 0.6$ nm is the monolayer edge of the MoS₂ flake. **(D)** Atomic resolution images from the areas indicated in **(C)**. The WSe₂ (left) is twisted with respect to the MoS₂ (right), as indicated by the dashed black lines tracing the atomic rows in each image. In addition, the moiré pattern is visible in the image of the MoS₂. **(E)** Series of spectroscopy taken at multiple points in the moiré pattern (like that in Fig. S1A and B). Virtually no band edge variation is observed. In addition, no peaks associated with quantum-confined states are observed.

Zooming into the different regions of the large scale STM images allows us to explicitly confirm the twist angle of the layers. Atomic resolution of the WSe₂ and MoS₂ are shown in Fig. S2D on the left and right, respectively. Dashed lines are drawn along one of the atomic rows in each image, highlighting the small twist angle of about 15°, consistent with the observed moiré pattern which can be seen in the MoS₂ atomic resolution images. As discussed in the main text, no confinement or spatial variation of spectroscopic features are observed. Fig. S2E shows a series of spectra taken along a line like that in Fig. S1A and B. Plotted on top of each other, it is evident that there is virtually no spatial variation of the spectra as well as no sharp peaks like those at the AB_W and AB_{Se} locations of the heterobilayer with $\theta = 0°$.

## Model of the Moiré Pattern Hamiltonian

As described in the main text, in a semi-classical picture, the band edge variation amounts to an effective potential with a periodicity of the moiré pattern. To describe a real valued scalar function $f(\vec{r})$ that has the periodicity of the moiré pattern, it is convenient to take the Fourier decomposition:

$$f(\vec{r}) = \sum_{i,j} f_{i,j} e^{i\vec{G}_{i,j}\cdot\vec{r}}$$

where $\vec{G}_{i,j}$ are reciprocal lattice vectors. The real space moiré pattern and the first shell of reciprocal lattice points are shown in Fig. S3A and B, respectively. Following the method of Ref. (4), for the case of a slowly varying function in real space, we can approximate the function considering only the terms within the first shell of reciprocal lattice vectors. The sum then goes over the origin and the six points labeled $G_{i,j}$ in Fig. S3B. The symmetry and fact that the function must be real reduces the number of relevant Fourier coefficients to only two: a real zeroth order component $f_0$ and a complex first order component $f_G = |f_G|e^{i\phi}$. The function can then be shown to be

$$f(x,y) \approx f_0 + 4|f_G|\cos\left(G_1\frac{\sqrt{3}}{2}x\right)\cos\left(G_1\frac{y}{2}+\phi\right) + 2|f_G|\cos(G_1 y - \phi)$$

where $G_1 = \frac{4\pi}{\sqrt{3}L}$ and the two Fourier coefficients are defined by the value of the function at the high symmetry points $f(0,0) = A, f\left(\frac{1}{2}L, \frac{1}{2\sqrt{3}}L\right) = B, f\left(0, \frac{1}{\sqrt{3}}L\right) = C$ shown in Fig. S3A:

$$f_0 = \frac{A+B+C}{3}, |f_G| = \frac{2A-B-C}{18\cos\phi}, \tan\phi = \frac{\sqrt{3}(B-C)}{2A-B-C}$$

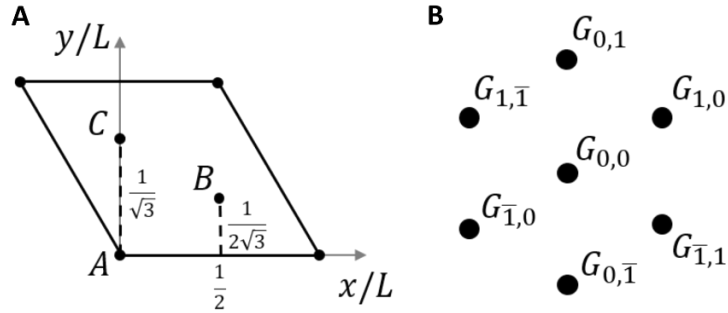

**Fig. S9. Fourier decomposition of the potential. (A)** Real space unit cell of the moiré pattern. A smoothly varying function $f(\vec{r})$ can be approximated by a simple first-order Fourier decomposition, where the Fourier coefficients are determined by the value of $f(\vec{r})$ at the high symmetry points $A, B,$ and $C$, indicated in the diagram. **(B)** The first shell of reciprocal lattice points for the unit cell shown in **(A)**. The distance between each set of neighboring points is $G_1 = \frac{4\pi}{\sqrt{3}L}$.

Using the above equations, we can find the Fourier coefficients describing the potential associated with a band edge.

$$V(\vec{r}) = \sum_{i,j} V_{i,j} e^{i\vec{G}_{i,j}\cdot\vec{r}}$$

For example, the $\Gamma_W$ band edge energy using the zero strain DFT results at the special spatial locations in the moiré pattern ($A = \Gamma_W^{AA} = -5.33$ eV, $B = \Gamma_W^{ABW} = -5.12$ eV, and $C = \Gamma_W^{ABSe} = -5.09$ eV) yields:

$$V_0 = -5.18 \text{ eV}, |V_G| = 0.025 \text{ eV}, \phi = -172.8°$$

The Hamiltonian is

$$H = -\frac{\hbar^2}{2m^*}\nabla^2 + V(\vec{r})$$

where we assume an effective mass of unity for simplicity, i.e. $m^* = m_e$. Taking advantage of the Fourier decomposition, the Hamiltonian can be expressed in a plane wave basis:

$$H = \begin{pmatrix} \ddots & V_G & 0 & 0 \\ V_G^* & \frac{\hbar^2|\vec{k}|^2}{2m^*} & V_G & 0 \\ 0 & V_G^* & \frac{\hbar^2|\vec{k}|^2}{2m^*} & V_G \\ 0 & 0 & V_G^* & \ddots \end{pmatrix}$$

where we have neglected the $V_0$ term, which amounts to a constant energy offset. This resembles a nearly free electron model mathematically, where the term $V_G$ opens a gap of size $2V_G$ at the Brillouin zone boundary. The band structures for $|V_G| = 0$ and $|V_G| = 0.025$ eV are shown in the main text in Fig. 2D.

The validity of this approach relies on a slow spatial variation of the band edge energy and an accurate value of the band edge energy from DFT calculations for the special high-symmetry points. Both of these conditions are violated to some degree for a large twist angle: any band edge energy variation would be over a much smaller spatial scale than for the zero twist angle case and with a twist angle, the DFT results are less accurate since the calculations have to match the lattice constants and necessitate a zero (or 180°) twist angle, as discussed below. In addition, one would expect that the individual layers would act more independently with a twist angle, i.e. less hybridization, deformation effects, etc., such that there would be a smaller spatial modulation of the band edge energy and thus a broader dispersion and no confinement. These conclusions are supported by the lack of band edge shifts and the lack of confinement observed in the case of the twisted heterobilayer, as shown in Fig. 2 and Fig. S2E.

## Strain Measurement

In order to obtain a direct measurement of the strain from our topographic images, we do a careful calibration of the STM piezo constants by measuring the lattice constant of epitaxial graphene and comparing that to the literature value. We repeat this calibration for multiple STM tip positions, where there are patches of graphene exposed. This allows us to make an absolute measurement of the atomic positions to extract lattice constants of the $MoS_2$ at each registry.

Since the measured corrugation of the heterostructure (~1.2 Å) and the thickness of the individual layers (~0.7 Å) are comparable, we investigated whether the strain is uniform along both lattice vectors and if there is any sheer strain. The components of the strain tensor are defined as

$$\varepsilon_{aa} = \frac{|\vec{a}| - a_0}{a_0}, \varepsilon_{bb} = \frac{|\vec{b}| - a_0}{a_0}, \varepsilon_{ab} = \frac{1}{2}\tan(\gamma - 60°)$$

where the axial strains $\varepsilon_{aa}$ and $\varepsilon_{bb}$ are defined to be positive when the lattice is stretched (tensile strain) and negative when the lattice is compressed (compressive strain) as compared to the nominal lattice of MoS$_2$ ($a_0 = 3.16$ Å). The lattice vectors $\vec{a}$ and $\vec{b}$ are shown in Fig. 3B of the main text. The sheer strain $\varepsilon_{ab}$ is defined in analogy to that for a rectangular unit cell, measuring the deviation from the ideal unit cell which has 60° interior angles, where $\gamma$ is the angle between the lattice vectors. To measure the strain for each registry, we utilize the Gwyddion 'Measure Lattice' feature which returns best fits of the two lattice vectors. This measurement is repeated multiple times for each registry, the results of which are shown in Fig. 3C of the main text. We find that for all registries, the sheer strain is consistent with zero, and that the two axial strains are consistent at each registry. The shear strains are found to be $(0.16 \pm 1.5)\%$, $(0.29 \pm 1.21)\%$, and $(-0.33 \pm 0.37)\%$ for the AA, AB$_W$, and AB$_{Se}$ registries, respectively. The AB$_W$ registry where $\varepsilon_{aa}$ found to be slightly lower than $\varepsilon_{bb}$. This result is surprising, and it is inconsistent with the the theoretical strain values reported in the main text. We tentatively interpret this difference between the apparent $\varepsilon_{aa}$ and $\varepsilon_{bb}$ strain values for the AB$_W$ registry as due to the lattice distortions arising from lateral forces between the tip and sample, for which we have some direct experimental evidence as presented below.

As mentioned in the main text, atomic resolution was achieved for small tunneling voltages, which means that the STM tip must be pushed in very close to maintain an appreciable tunneling current. Therefore, lateral forces between the tip and sample could have a dominant effect and influence the observed strain signal. As we discuss below, the moiré pattern is formed when domains of different registries are formed. If there are large lateral forces due to the presence of the tip, then one would expect to see changes of the moiré domains for those conditions with rather extreme tunneling conditions. The atomic resolution images might be one example of this, but a clearer example occurs at even smaller sample voltages. Figure S5 shows three STM images of the same area, all taken with the same setpoint current. The leftmost image shows the familiar moiré pattern. At very small values of $V_S$, we see a very different 'checkerboard' pattern form (also observed, though less distinctly, in Ref. [2]), but still with the same periodicity as before. At these tunneling conditions, both for small positive and negative sample biases, we see these two distinct domains forming. What is likely happening here is that lateral forces due to the presence of the tip are pushing on the surface, to the point where it becomes energetically favorable for the top layer to be pushed into one of the two the lower energy configurations such that a sort of point-like stacking fault is created at the AA sites (bright spots in the +1.5 V image). Therefore, when the tip is pressed up close to the sample, we are seeing the layers being 'locked' into separate domains of the AB$_W$ and AB$_{Se}$ registries.

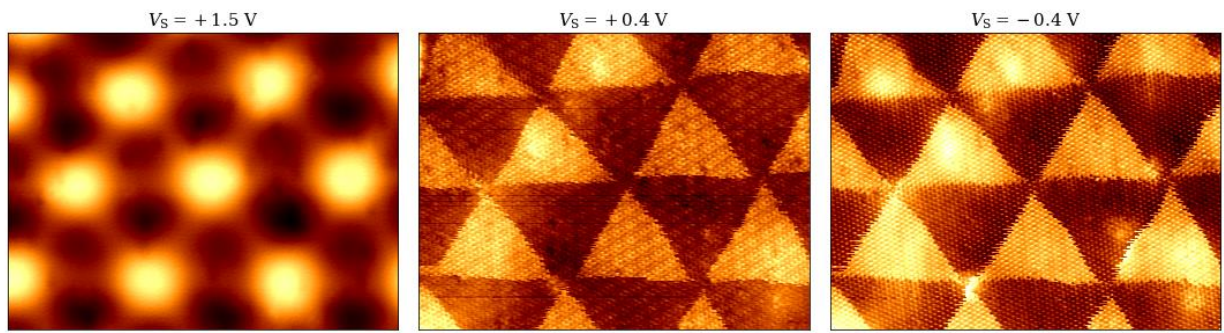

**Fig. S10. Electrostatic force effects on observed topographic signals.** STM topographic images for the same area and same setpoint current, but different sample voltages, indicated above each panel. The small tunneling voltage requires the STM tip to be very close to the sample for the measurement at ±0.4 V images, such that atoms lock into the two low energy stacking configurations. The color scale is the same for each image.

## Theoretically Predicted Strain Variation

Following Ref. (*5*), we assume the relaxation of the moiré superlattice can be described by two displacement fields, $u^t$ for the MoS$_2$ and $u^b$ for WSe$_2$. These two fields completely specify the local registries in the moiré superlattice, and the interlayer potential energy is obtained by summing over energies for these local registries. The stiffness tensors for both MoS$_2$ and WSe$_2$ are assumed to be rotationally symmetric and are described by bulk and shear modulus. The bulk moduli for MoS$_2$ and WSe$_2$ was found to be 45,235 meV/unit cell and 42,679 meV/unit cell, respectively. The shear moduli for MoS$_2$ and WSe$_2$ were found to be 27,186 meV/unit cell and 28,965 meV/unit cell, respectively. We numerically discretize the two displacement fields in the real space and minimize the total energy by conjugate gradient descent, as implemented in the Julia package (*6*). In the minimization, three-fold rotational symmetry is enforced to reduce the number of variables.

The interlayer energy after relaxation is plotted in Fig. 56, where AA, AB$_W$ and AB$_{Se}$ registries can be easily recognized. The strain values for different registries are obtained by taking the divergence of $u^t$ and $u^b$. The strains for the AA, AB$_W$, and AB$_{Se}$ registries were found to be -1.46%, 0.88%, and 1.14% in the MoS$_2$ layer and 1.60%, -0.96%, and -1.24% in the WSe$_2$ layer, respectively.

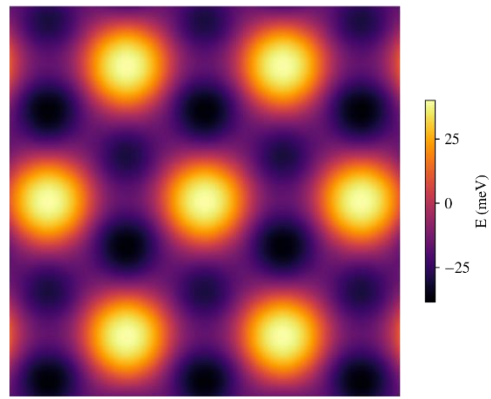

## Interpreting the Measured Corrugation from Topography

**Figure S11. Interlayer potential energy per unit cell in real space.** The bright patches correspond to the AA registry. The darker patches are the AB$_W$ and AB$_{Se}$ registries. The AB$_{Se}$ registry has a slightly lower value of interlayer potential energy, since three-fold rotational symmetry was enforced (rather than six-fold rotational symmetry). See text for details.

Any constant-current measurement in an STM experiment requires great care if it is to be interpreted as a physical topography (i.e. atomic heights) of the surface. In general, different tunneling parameters will yield different constant-current measurements, as demonstrated for this sample in the voltage dependent results shown in the Supporting Information of Ref. (*1*). Therefore, identifying the measured corrugation height at large positive voltage as the *true,* physical corrugation requires justification. Figure S6A shows constant-current profiles acquired in two different measurements, at $V_S = +1.5$ V and $V_S = -1.5$ V, from which it can be seen that the ordering of the two corrugation minima swap. At large positive sample voltage, the AB$_{Se}$ location is the absolute minima of the corrugation. At large negative sample voltage, the AB$_W$ location is the absolute minima. The difference between the two measured signals can be attributed to the complicated electronic structure of the heterostructure.

To identify which ordering is correct, i.e. which location is the actual corrugation minima, we must turn to spectroscopic results. In Fig. S6B, we show representative spectra at each registry, along with a voltage dependent noise level (*1*). The differential conductance measured in STS is, as a first order approximation, proportional to the local density of states. We see that at large negative sample voltages there are multiple bands that contribute to the spectra, the $\Gamma_W$, $K_M$, and $\Gamma_{W2}$ bands. The onsets of these bands vary depending on spatial location, which makes interpretation of these constant-current measurements in terms of physical topography difficult. On the other hand, at large positive sample voltages, there is predominantly a single band that contributes, the $Q_M$ band. Spectroscopic onsets for this band are relatively independent of voltage, making a direct comparison between registries more straight-forward. In addition, the tunneling current in STM measurements is typically dominated by higher lying states in the sample. For negative sample bias, the higher lying states are those corresponding to the onset of the band edge, such that the measurement is very sensitive to the position of the band edge. For positive sample bias, the higher lying states are those higher in energy than the onset of the band edge, such that the measurement is less sensitive to the position of the band edge. We therefore take the large positive sample voltage ($V_S = +1.5$ V or $V_S = +2.0$ V) as representative of the true ordering of the corrugation heights. This conclusion aligns with our strain measurement as well.

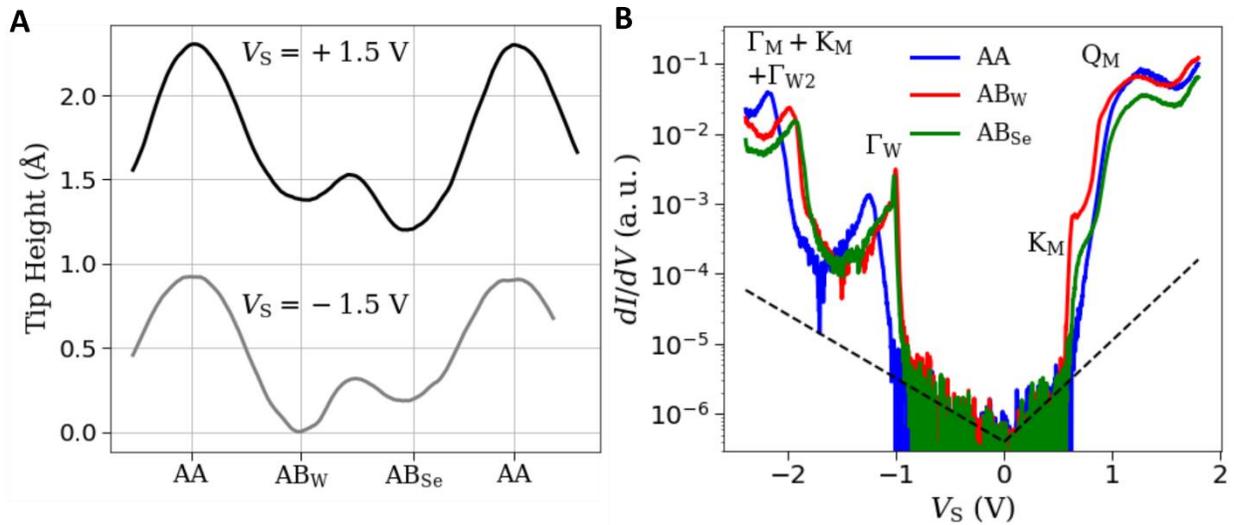

**Fig. S12. Interpreting topographic measurements of the heterostructure. (A)** Profiles taken from STM topographs at different sample voltages. Notably, the absolute minima of the corrugation swap when going from large positive to large negative sample voltage, due to the complicated electronic structure of the heterobilayer. **(B)** Representative spectra taken at all three locations, with a voltage dependent noise level shown as the black dashed line. The complicated band shifts due to atomic registry and mechanical deformations make it difficult to attribute topographic measurements at large negative voltages as true height measurements, while variations are less prominent at large positive voltages. See text for details.

## Band Structure Calculations

Density functional theory (DFT) calculations were performed to investigate the dependence of the electronic structure on mechanical deformations. For the heterostructure, calculations of 1x1 unit cells of $MoS_2$ on $WSe_2$ were computed for each registry. We assume a zero degree alignment (as opposed to 180 degree, as detailed in Ref (*1*)). To compare band edge energies between different registries in the heterostructure, we take the reference to be the vacuum level of the $WSe_2$ ($E_{vac}$). As described in the next section, we also perform calculations for the individual monolayers as a function of strain. For the monolayer calculations, we take the vacuum level of the respective monolayer as the reference.

For the strain dependent calculations, the structure is allowed to relax while holding the lattice constants of the two materials fixed. For simplicity of our calculations, we choose to assume uniform biaxial strain ($\varepsilon = \varepsilon_{aa} = \varepsilon_{bb}, \varepsilon_{ab} = 0$) in both layers of the heterostructure. For the separation dependent calculations, the structure is fixed to the average lattice constant of the two materials $a = \bar{a} = 3.22$ Å and the interlayer separation is fixed at the reported values. The optimal interlayer separation reported in the main text, $d_0 = 6.9$, 6.32, and 6.33 Å for the AA, AB$_W$, and AB$_{Se}$ locations, respectively, is determined by setting the lattice constant of each layer to the average lattice constant and minimizing the energy of the structure.

## Deformation Potentials

DFT calculations were performed for isolated monolayers of WSe$_2$ and MoS$_2$, as well as the heterostructure for each registry. To separate the effects of mechanical deformations and hybridization, we examine the position of the band edges at special points in the Brillouin zone as a function of lattice constant for each system. The results are shown in Fig. S7 for the AA registry and monolayers of WSe$_2$ and MoS$_2$. It can be seen that the valence (conduction) band of the heterostructure primarily derives from WSe$_2$ (MoS$_2$) states, as reported previously. With these calculations done, we track the band edges as we vary the lattice constant in each calculation as shown.

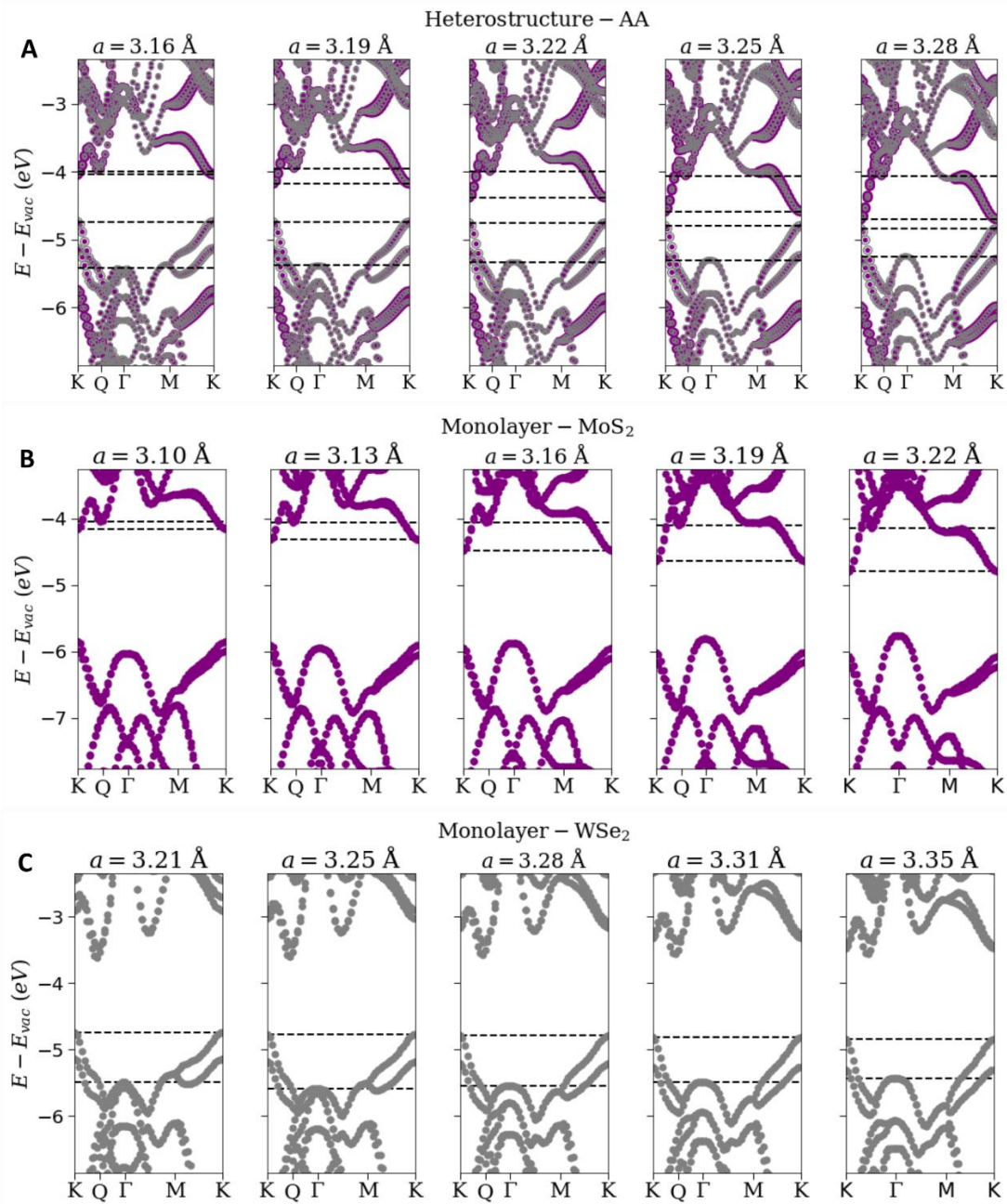

**Fig. S13. DFT band structure calculations as a function of lattice constant.** DFT calculations for different values of the lattice constant for the **(A)** heterostructure AA registry **(B)** monolayer MoS$_2$ and **(C)** monolayer WSe$_2$. The K$_M$ and Q$_M$ conduction band edges are marked by dashed lines in (A) and (B) and the Γ$_W$ and K$_W$ valence band edges are marked in (A) and (C).

First, let us turn our attention to the conduction band (i.e. MoS$_2$ derived states). If we compare the heterostructure calculation to the MoS$_2$ calculation at the same value of the lattice constant, we see that the difference between the Q$_M$ and K$_M$ band edge energies are different ($\Delta E_{Q,K} = 0.38$ eV for the heterostructure and $\Delta E_{Q,K} = 0.65$ eV for the monolayer MoS$_2$ with $a = \bar{a} = 3.22$ Å). However, as we vary the lattice constant, the change in the band edges are very similar. This is demonstrated in Table S1, where we show linear fits of the band edges as a function of strain, i.e.

$$E = \phi\varepsilon + E_0$$

where $\phi$ is the deformation potential and $E_0$ is the zero-strain position of the band edge, relative to the vacuum level of the system. We see that the extracted deformation potentials for the conduction band edges are nearly the same for the heterostructure and monolayer $MoS_2$. We therefore assume that shifts in the absolute position of the $K_M$ and $Q_M$ band edges can be associated with strain in the $MoS_2$. The effects of hybridization are then accounted for in the zero-strain position of the band edge.

**Table S1. Strain deformation potentials from DFT calculations.** Deformation potentials in meV/% strain. For the heterobilayer, the conduction band deformation potentials ($Q_M$ and $K_M$ band edges) are associated with strain in the $MoS_2$ and the valence band deformation potentials ($K_W$ and $\Gamma_W$) are associated with strain in the $WSe_2$. Asterisk on the $\Gamma_W$ monolayer deformation potential to highlight that we only consider the calculations for lattice constants $a = 3.25$ Å and higher, as explained in the text.

|  | Monolayer |  | Heterobilayer |  |  |  |
| --- | --- | --- | --- | --- | --- | --- |
|  | $MoS_2$ | $WSe_2$ | AA | $AB_W$ | $AB_{Se}$ | Mean |
| $Q_M$ | $-2.2 \pm 0.5$ | N/A | $-2.4 \pm 0.9$ | $-2.0 \pm 0.9$ | $-1.9 \pm 0.9$ | $-2.1 \pm 0.5$ |
| $K_M$ | $-16.0 \pm 0.1$ | N/A | $-16.9 \pm 0.9$ | $-16.5 \pm 0.4$ | $-16.9 \pm 0.8$ | $-16.8 \pm 0.4$ |
| $K_W$ | N/A | $-2.6 \pm 0.2$ | $-2.5 \pm 0.4$ | $-2.3 \pm 0.3$ | $-2.8 \pm 0.4$ | $-2.5 \pm 0.2$ |
| $\Gamma_W$ | N/A | $5.2 \pm 0.3*$ | $4.4 \pm 0.2$ | $4.7 \pm 0.6$ | $3.8 \pm 0.3$ | $4.3 \pm 0.2$ |

Now, turning to the valence band (i.e. $WSe_2$ derived states), we can repeat the same procedure. We see that associating the changes in the conduction band with strain in the $WSe_2$ again aligns well for the $K_W$ band, but the $\Gamma_W$ band edge is more complicated. Particularly, there is a lower lying band which creeps upward for large compressive strains and destroys the linearity in motion of the band edge with strain. However, in the heterostructure, these bands are still evident, but further apart due to the hybridization effects. The two $\Gamma_W$ bands therefore only intersect at very large values of compressive strain for the $WSe_2$. As discussed in the main text, since the $MoS_2$ is in maximal tensile strain when the layers are closest together, the $MoS_2$ 'compensates' such that when the layers are furthest apart the $MoS_2$ is compressed. In light of the evidence that there is in-plane and out-of-plane deformations in the $WSe_2$ as well, we expect the opposite behavior for the $WSe_2$ (i.e. the maximum *compressive* strain would be achieved when the layers are closest together to match the lattice constants and the maximum *tensile* strain when they are furthers apart). Therefore, we throw out the monolayer $WSe_2$ calculation for the largest compressive strain when determining the deformation potential and find good agreement between the monolayer and heterostructure calculations as we did in the case of the $MoS_2$ derived states.

As discussed in the main text, there is an additional effect on the electronic structure due to vertical separation between the layers of the heterostructure. We repeat the same procedure outlined above, but with the vertical separation being varied while fixing the lattice constant of the two layers to the average value of the lattice constant. We find that the $\Gamma_W$ band edge varies significantly with the separation of the layers (Fig. 5A) and that the $\Gamma_W$ band edge energy variation is well described by a third order polynomial (Fig. 5B). The $\Gamma_W$ band edge for each registry is fit to a polynomial of the form:

$$\Gamma_W = \bar{\Gamma}_{W,0} + f_1(d - d_0) + f_2(d - d_0)^2 + f_3(d - d_0)^3$$

Where $\bar{\Gamma}_{W,0}$ is the band edge position for the equilibrium spacing (i.e. the spacing determined by minimizing the energy while the separation distance is relaxed, and the lattice constant is fixed to the average value). We then compare the band edges for each registry with the value for the equilibrium spacing and fit the band edges to find the coefficients of the polynomial function. The results for each registry are shown in Table S2. There are changes at other points in the Brillouin zone, such as the $K_M$ band edge, but the changes are smaller and don't evolve as uniformly as the $\Gamma_W$ band edge.

**Table S2. Separation deformation potentials for the $\Gamma_W$ band edge.** Optimal separation distance, optimal separation band edge energy (relative to the vacuum level), and best fits to the polynomial function for each registry.

|   | AA | AB$_W$ | AB$_{Se}$ |
|---|---|---|---|
| $d_0$ (Å) | 6.90 | 6.32 | 6.33 |
| $\bar{\Gamma}_{W,0}$ (eV) | $-5.34 \pm 0.01$ | $-5.11 \pm 0.01$ | $-5.07 \pm 0.01$ |
| $f_1$ (eV/Å) | $-0.42 \pm 0.02$ | $-0.66 \pm 0.02$ | $-0.75 \pm 0.01$ |
| $f_2$ (eV/Å$^2$) | $0.45 \pm 0.01$ | $0.41 \pm 0.01$ | $0.45 \pm 0.01$ |
| $f_3$ (eV/Å$^3$) | $-0.20 \pm 0.02$ | $-0.09 \pm 0.02$ | $-0.09 \pm 0.02$ |

Note that while we compare corresponding band edges at different registries, we don't make a quantitative comparison between the valence and conduction band. It is well known that DFT underpredicts the value of the band gap, so comparing the $K_M$ and $\Gamma_W$ band edges would not result in a meaningful measure of the effects of mechanical deformations. The results for the obtained deformation potentials would also be sensitive to the details of the DFT calculations, but given the agreement between our deformation potentials with those obtained by experimental studies, we believe the evolution of band edges with mechanical deformations is largely captured by our theoretical analysis.

### Inferred WSe$_2$ Corrugation

In order to extract an inferred corrugation for the WSe$_2$, we take advantage of the deformation potentials obtained through DFT. Since the $\Gamma_W$ band edge is sensitive to the separation of the layers, we can use the observed band edges in STS to back out the corrugation of the WSe$_2$. Considering both strain and separation dependence, we can write the band edge energy at each location as

$$\Gamma_W(a, d) = \bar{\Gamma}_{W,0} + \phi(a - \bar{a}) + f_1(d - d_0) + f_2(d - d_0)^2 + f_3(d - d_0)^3$$

where $\phi$ is the deformation potential. From STS, we have two relevant measurements, the differences between the $\Gamma_W$ band edge energies i.e. $[\Gamma_W^{AA} - \Gamma_W^{AB_{Se}}]$ and $[\Gamma_W^{AB_W} - \Gamma_W^{AB_{Se}}]$. The deformation and separation potentials are all known from the fits to DFT results, leaving the lattice constants in the WSe$_2$ and separations between the layers to be determined.

For each registry, we use the experimentally determined energy difference of the $\Gamma_W$ band edge to extract the layer separation as a function of lattice constant. The height of the WSe$_2$ layer at each registry ($z_W$) can be inferred from the layer separation and the measured height of the MoS$_2$ ($z_M$), i.e. $z_W = z_M - d$. In addition, we can observe the step height of an MoS$_2$ island directly from STM (see Fig.

1 of Ref. (*1*)) and relate that to the optimal spacing found in DFT. In Fig. S8, we show the results for the inferred corrugation of the WSe$_2$ layer, relative to the height at the AB$_{Se}$ registry, as a function of the interlayer separation at the AB$_{Se}$ location. The black dashed line shows the measured step height in STM, with the grey dashed region indicating the standard deviation of that measurement. We see that within the range of experimentally determined separation at the AB$_{Se}$ registry, the WSe$_2$ corrugation is in phase with the MoS$_2$ corrugation that is directly observed (indicated by the blue and red lines being larger than zero in the grey dashed region). The magnitude of the corrugation in the WSe$_2$ varies based on the value of the experimentally determined step height, but these results indicate that the WSe$_2$ corrugation is indeed in phase with the MoS$_2$ corrugation.

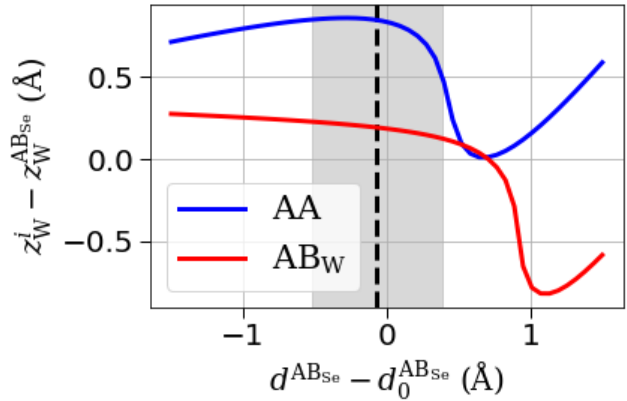

**Figure S14. Inferred corrugation of the WSe$_2$ layer.** Height difference of the WSe$_2$ layer between the AA (AB$_W$) and AB$_{Se}$ locations in blue (red), as a function of interlayer separation for the AB$_{Se}$ location. Dashed black line and the grey shaded region indicate the inferred separation at the AB$_{Se}$ registry as determined from the step height observed in STM and the standard deviation of that measurement, respectively.